\documentclass[aps,prb,groupedaddress,twocolumn]{revtex4}
\usepackage{epsfig}
\usepackage[dvipsnames,usenames]{color}
\usepackage{color}
\usepackage{amsmath}
\usepackage{comment}
\usepackage{array}
\usepackage{multirow}
\usepackage{tabularx}
\usepackage{float}
\usepackage{hyperref}

\emergencystretch=\maxdimen
\begin{document}
\title{Enhanced tendency of $d$-wave pairing and antiferromagnetism \\ in doped staggered periodic Anderson model}
\author{Mi Jiang}
\affiliation{Stewart Blusson Quantum Matter Institute, University of British Columbia, Vancouver, BC, Canada}

\begin{abstract}
Whether or not a physical property can be enhanced in an inhomogeneous system compared with its homogeneous counterpart is an intriguing fundamental question. We provide a concrete example with positive answer by uncovering a remarkable enhancement of both antiferromagnetic (AF) structure factor and $d$-wave pairing tendency in the doped staggered periodic Anderson model (PAM) with two alternating inequivalent local moments. The common thread of these enhancement is found to originate from the generic ``self-averaging'' effect and non-monotonic dependence of the corresponding physical quantity in homogeneous PAM. More strikingly, we provided evidence of the coexistence of these two enhancement via a tentative phase diagram. Our findings may imply the plausible generalization of enhancing physical properties in generic inhomogeneous systems.
\end{abstract}

\maketitle

\section{Introduction}
Inhomogeneous systems are ubiquitous amongst both real materials and artificial lattice/superlattices. 
One fundamental question is the effects of the inhomogeneity on the physical properties in originally homogeneous systems. In particular, whether or not a desired property can be enhanced in the presence of inhomogeneity is a fascinating problem of broad interest.
One particularly important example of inhomogeneous systems is the family of heavy fermion compounds consisting of two or more inequivalent local moment sites per unit cell~\cite{ternary1,ternary2,ternary3,ternary4,ternary4a,ternary5,ternary5a,ternary6,ternary7,ternary8,Custers2020}, which has been uncovered to possess a wide range of intriguing phenomena, for instance, two successive magnetic phase transitions in (RE)$_3$Pd$_{20}$X$_6$ (RE = rare earth, X = Si, Ge)~\cite{ternary1,ternary2,ternary3,ternary4,ternary4a}, intermediate valence in Ce-Ru-X (X = In, Al, Sn)~\cite{ternary5,ternary5a}, coexistence of antiferromagnetism and superconductivity in Ce$_3$(Pt,Pd)In$_{11}$~\cite{ternary6,ternary7,ternary8,Custers2020}, due to the interplay between multiple energy scales from inequivalent $f$ local moments, which is absent in systems consisting of only one rare-earth ion per unit cell.
Undoubtedly, these recent experimental findings call for theoretical understanding of the inhomogeneity effects due to the additional rare-earth ion. 

On the one hand, the Kondo/Anderson lattice models (KLM/PAM) on a square lattice is conventionally believed to capture the essential physics of heavy-fermion materials via a lattice of two orbitals of electrons, one of which is strongly correlated $f$ electrons that hybridize with the conduction electrons~\cite{pam1,pam2,pam3,pam4,pam5,PAMreference}. In essence, KLM/PAM describes the competition between the phase of antiferromagnetism between $f$ local moments induced via the indirect Ruderman-Kittel-Kasuya-Yosida (RKKY) interaction~\cite{PAMRTS} mediated by the conduction band at small $c-f$ hybridization and the phase of paramagnetic spin-liquid ground state of independent Kondo singlets formed by strong hybridization between the conduction and local $f$ electrons. 

On the other hand, two or more inequivalent rare-earth ions can be mimicked by KLM/PAM with multiple correlated $f$ orbitals with distinct hybridization with the conduction electrons in a single unit cell in order to  
explore whether the distinct Kondo screening energy scales can generate some new physics. 
In this approach, the simplest staggered PAM (sPAM) with two alternating inequivalent local moments has been studied to demonstrate the competitive versus cooperative Kondo screening~\cite{Vojta2011} and the existence of a compressible ferrimagnetic phase~\cite{Costa2018}. 
These previous work have provided a great deal of insights on the rich physics induced by the additional inhomogeneity compared with the conventional homogeneous PAM. 
Recently, a striking enhancement of antiferromagnetism of $f$ local moments has been discovered in sPAM at half-filling~\cite{MJ2020}.
This previous work confirmed the expectation that the lattice antiferromagnetism can be remarkably enhanced if the hybridization strengths of two inequivalent local moments with the conduction electrons reside in the Kondo singlet and antiferromagnetism regime separately of the phase diagram of the homogeneous PAM. In other words, the antiferromagnetic correlation can be largely enhanced via replacing a sublattice of local moments by singlets with much stronger $c-f$ hybridization. This counterintuitively new route of strengthening the AF correlations can be interpreted as the consequence of the (a) ``self-averaging'' effect between two distinct local moments together with (b) the non-monotonic dependence of AF correlations in homogeneous PAM.
More importantly, we pointed out that these two essential ingredients strongly suggests the possible enhancement of physical properties in other systems, for example, the Hubbard model with staggered interaction strengths due to its underlying connection with PAM~\cite{note0,Held2000,yifeng}.

Nonetheless, the previous discovery is only limited to the case of half-filled orbitals. It has been well established that the doped conventional PAM can feature the competition and coexistence between the antiferromagnetism and $d$-wave superconductivity~\cite{WeiWu}. Therefore, the doping effect on the enhanced AF and accordingly its competition and/or coexistence with the superconductivity in sPAM, which is naturally a fundamental model relevant to the family of heavy fermion compounds~\cite{ternary6,ternary7,ternary8}, is a fascinating question to answer. To this aim, here we extend our previous study~\cite{MJ2020} to the doped system. Specifically, we will demonstrate that the enhanced ordering tendency of the antiferromagnetism can persist in a broad range of doping levels until a critical density. Besides, we will provide evidence that the $d$-wave pairing tendency can also be enhanced via the same mechanism.  
More strikingly, our tentative phase diagram at a particular density shows the possible coexistence of these two enhancement.

The paper is organized as follows: Section II discusses
the staggered PAM Hamiltonian and the definitions of some key physical quantities calculated.
Section III and IV demonstrate the enhancement of antiferromagnetism and $d$-wave pairing tendency separately. Section V discusses the competition and coexistence between their enhancement via a tentative phase diagram. Section VI summarizes our results.

\section{Model and methodology}
We adopt the same sPAM with two alternating inequivalent local moments on two-dimensional square lattice as our previous work~\cite{MJ2020}, which reads in the half-filled form:
\begin{eqnarray}
    {\cal H} = &-& t \sum\limits_{\langle ij \rangle \sigma}
(c^{\dagger}_{i\sigma}c_{j\sigma}^{\vphantom{dagger}}
+c^{\dagger}_{j\sigma}c_{i\sigma}^{\vphantom{dagger}}) 
+ \sum\limits_{i \sigma}  V_i (c^{\dagger}_{i\sigma}f_{i\sigma}^{\vphantom{dagger}}+ f^{\dagger}_{i\sigma}c_{i\sigma}^{\vphantom{dagger}}) \nonumber \\
    &+& U \sum\limits_{i} (n^{f}_{i\uparrow}-\frac{1}{2}) (n^{f}_{i\downarrow}-\frac{1}{2})
- \mu \sum\limits_{i\sigma} (n^{c}_{i\sigma}+ n^{f}_{i\sigma} )
\label{inPAM}
\end{eqnarray}
where $c^{\dagger}_{i\sigma}(c_{i\sigma}^{\vphantom{dagger}})$
and $f^{\dagger}_{i\sigma}(f_{i\sigma}^{\vphantom{dagger}})$
are creation(destruction) operators for conduction and local electrons on site $i$ with spin $\sigma$.
$n^{c,f}_{i\sigma}$ are the associated number operators.
Here the chemical potential can be tuned to reach a desired average occupancy of the lattice. In particular, $\mu=0$ corresponds to the case that both $c$ and $f$ orbitals are individually half-filled studied previously~\cite{MJ2020}.
The hopping $t=1$ between conduction electrons on
nearest neighbor sites $\langle ij \rangle$ sets the energy scale. $U$ is the local repulsive interaction in the f orbital. The sublattice dependent $V_i=V_1, V_2$ are two distinct hybridizations between conduction electrons and two inequivalent local moments respectively.
In the absence of Hubbard interaction, the two-site two-orbital unit cell gives rise to four energy bands such that the system is readily a band insulator for any finite $V_{1,2}$ at particular fillings, e.g. half-filling and one quarter filling. See more details on the non-interacting case in Appendix.~\ref{U0}. In addition, we use the same numerical technique of finite temperature determinant Quantum Monte Carlo (DQMC) as Ref.~\cite{MJ2020} to explore the physics of doped staggered PAM. As a celebrated computational method, we refer the readers for detailed discussion of the formalism to Ref.~\cite{blankenbecler81}. Appendix.~\ref{dqmc} also provides some brief introduction.

Throughout the paper, we assume that $V_1 \geq V_2$ and concentrate on the characteristic Hubbard interaction $U=4.0t$ as before~\cite{MJ2020}. Besides, we treat both conduction and $f$ orbitals as lattice sites so that the reported density $\rho$ is physically average density of two orbitals.
Due to the infamous fermionic sign problem in the doped sPAM (see Appendix. B), we have to utilize a smaller lattice size $8 \times 8$ in most cases instead of $12 \times 12$ used at half-filling to study the physics at low enough temperature. 
To explore the possibility of AF enhancement in the doped sPAM, we calculate the antiferromagnetic (AF) structure factor of f-orbital local moments 
\begin{equation}
S^f_{AF}(V_1,V_2)=\frac{1}{N} \sum_{ij} e^{-i \mathbf{q} \cdot (\mathbf{R_i}-\mathbf{R_j})} \langle (n^f_{i\uparrow}-n^f_{i\downarrow}) (n^f_{j\uparrow}-n^f_{j\downarrow}) \rangle
\end{equation}
at $\mathbf{q}=(\pi,\pi)$, where $\mathbf{R_i}$ denotes the coordinates of site $i$ and $N$ is the lattice size. 

In addition, as mentioned before, the doped PAM can host the $d$-wave superconductivity~\cite{WeiWu}. Therefore, it is intriguing  to investigate the interplay between the enhanced AF and $d$-wave superconductivity via calculating the $d$-wave pairing susceptibility of $f$-electrons
\begin{eqnarray}
P_d &=& \frac{1}{N}\frac{1}{G} \sum_{ij} \sum_{\delta \delta'} g(\delta')g^*(\delta) \nonumber \\
&\times & \int^{\beta}_0  \langle f_{j+\delta',\downarrow}(\tau) f_{j,\uparrow}(\tau) f^{\dagger}_{i,\uparrow}(0) f^{\dagger}_{i+\delta,\downarrow}(0) \rangle \  d\tau
\label{Pd}
\end{eqnarray}
where $g(\delta)$ is the general form factor in real space and $G=\sum_{\delta} |g(\delta)|^2$ is the normalization factor. In particular, for $d$-wave pairing channel, $g(\delta)=\pm 1$ for nearest-neighbor separation $\delta$ along $x$ and $y$ directions respectively.
Similarly, we can define and calculate the uncorrelated
susceptibility $P^0_d$ that is the bubble contribution without vertex corrections~\cite{RTS1989,Scalapinobook}. In a DQMC simulation, the distinction between $P_d$ and $P^0_d$ lies in the evaluation of the expectation value of the four fermion terms in Eq.~\ref{Pd}. Precisely, $P_d$ ($P^0_d$) involves the Green's functions obtained by the Wick contractions first multiplied (averaged) and then averaged (multiplied).
Therefore, $P_d$ includes all interaction effects while $P^0_d$  only considers the interaction in the single-particle level.
We can further define the interaction vertex via
\begin{eqnarray}
\Gamma_d &=& \frac{1}{P_d} - \frac{1}{P^0_d}
\label{Gd1}
\end{eqnarray}
so that the sign of  $\Gamma_d P^0_d$ reflects whether the pairing interaction is repulsive (positive) or attractive
(negative). Eq.~\ref{Gd1} can be rewritten as
\begin{eqnarray}
P_d &=& \frac{P^0_d}{1+\Gamma^{\phantom{0}}_d P^0_d} 
\label{Gd}
\end{eqnarray}
so that $\Gamma^{\phantom{0}}_d P^0_d=-1$ signals the $d$-wave superconducting instability.

\section{AF enhancement}
\begin{figure}[h!] 
\psfig{figure=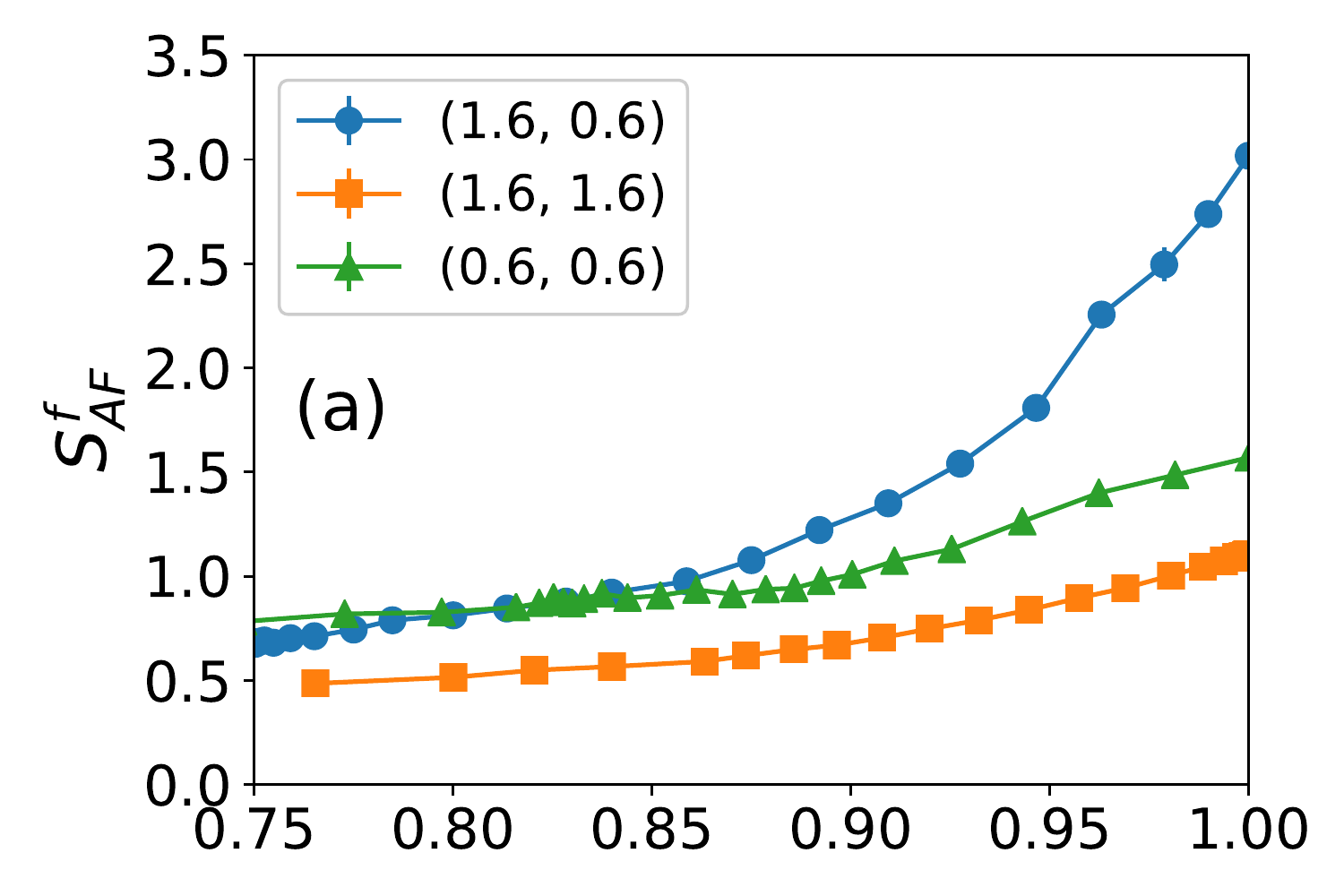,
width=.5\textwidth,angle=0,clip=true, trim = 0.0cm 0.5cm 0.0cm 0.2cm} \\
\psfig{figure=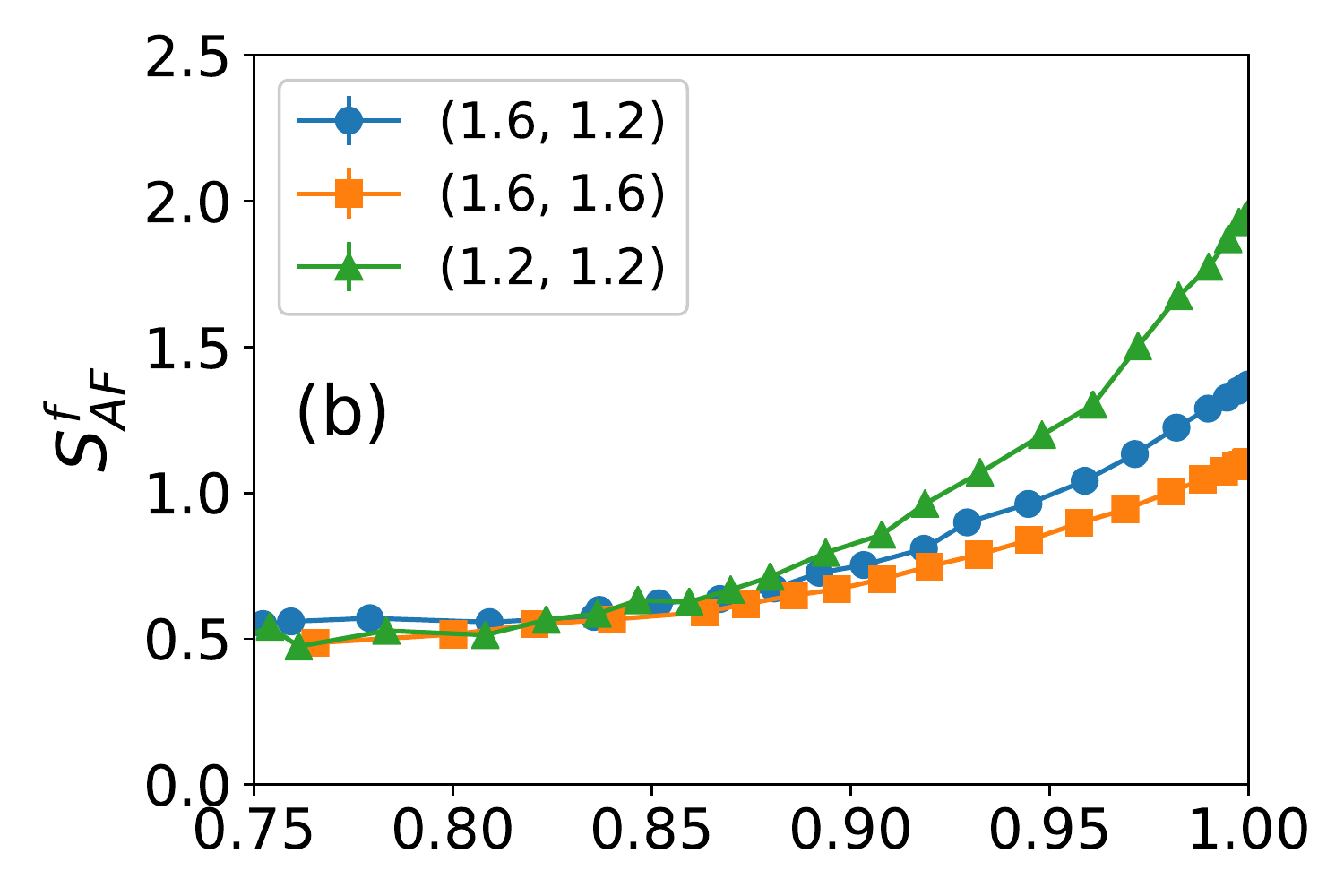,
width=.5\textwidth,angle=0,clip=true, trim = 0.0cm 0.5cm 0.0cm 0.2cm} \\
\psfig{figure=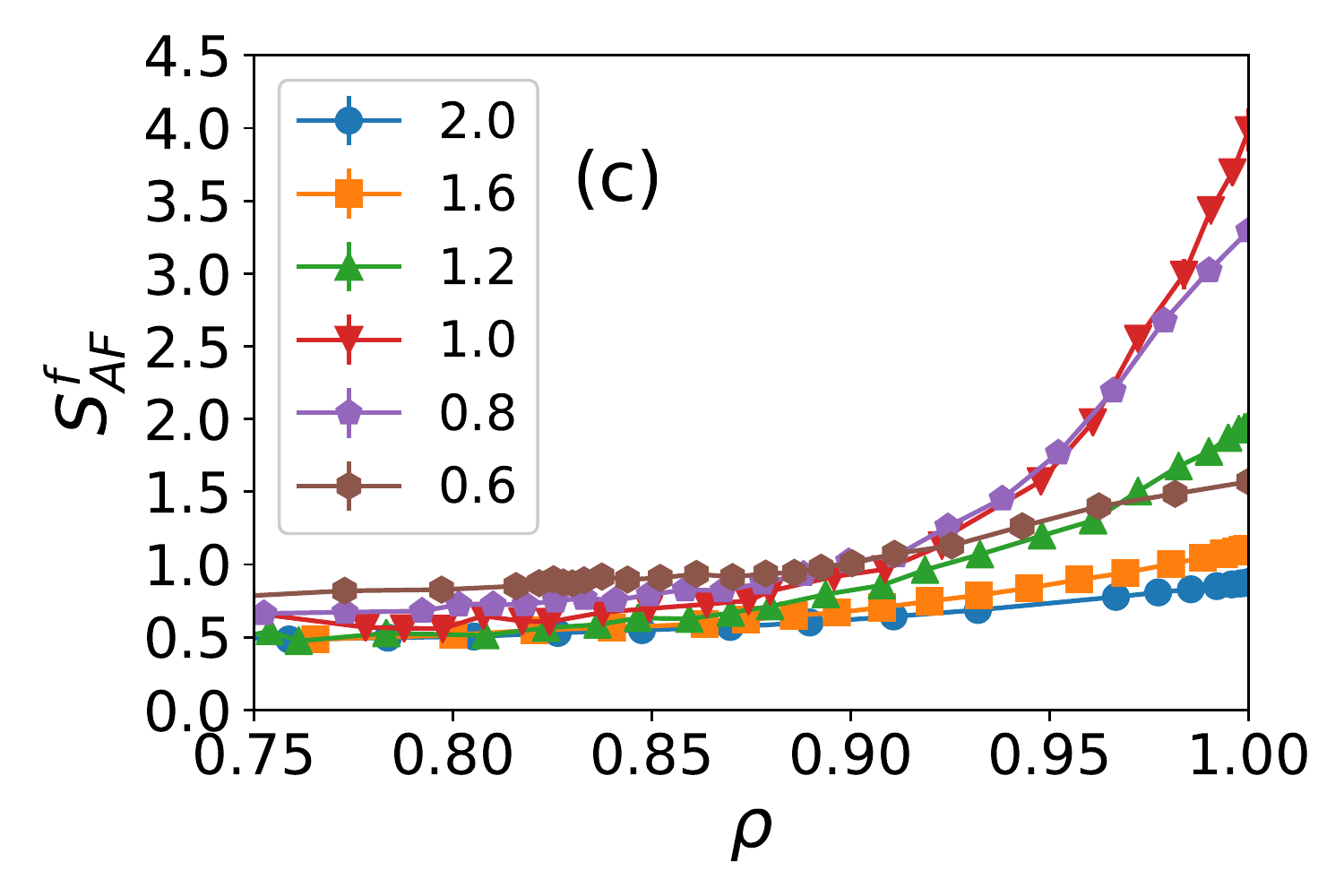,
width=.5\textwidth,angle=0,clip} 
\caption{(Color online) Doping evolution of $S^f_{AF}(V_1,V_2)$ at $\beta t=15$ in (a-b) two characteristic systems with distinct behavior compared with their counterparts in homogeneous PAM and (c) homogeneous PAM. The legend is shown as $(V_1,V_2)$ of sPAM in (a-b) and $V$ of PAM in (c).}
\label{Safden}
\end{figure}

We first illustrate our findings of the doping effects on the enhanced AF tendency at half-filling~\cite{MJ2020}. Figure~\ref{Safden} (a-b) show the doping evolution of $S^f_{AF}(V_1,V_2)$ of two characteristic systems with distinct behavior compared with their homogeneous counterparts; Fig.~\ref{Safden} (c) provides the referenced evolution of $S^f_{AF}(V)$ in homogeneous PAM. 
At half-filling $\rho=1$~\cite{MJ2020}, we reiterate that if $V_{1,2}$ both exceed $V_c \sim 1.0t$ or not of homogeneous PAM, namely both in the Kondo singlet or AF insulator regime, $S^f_{AF}(V_1)<S^f_{AF}(V_1,V_2)<S^f_{AF}(V_2)$. In contrast, if $V_{1,2}$ depart from each other and locate on the Kondo singlet or AF insulator regime separately, the distinct AF enhancement appears due to the combined effects of ``self-averaging'' and non-monotonic dependence of $S^f_{AF}(V)$ in homogeneous PAM. These features are supported by Fig.~\ref{Safden} despite the quantitative difference from that reported previously~\cite{MJ2020} due to the slightly higher temperature $\beta t=15$ at smaller lattice.

As expected, the doping destroys the AF evidenced by the gradual reduction of $S^f_{AF}$ in all cases. In addition, the AF enhancement of $S^f_{AF}(V_1,V_2)$ (Fig.~\ref{Safden}(a)) persists to a wide range of doping levels.
Conversely, in the case shown in Fig.~\ref{Safden}(b), the ``self-averaging'' induced $S^f_{AF}(V_1)<S^f_{AF}(V_1,V_2)<S^f_{AF}(V_2)$ is clear in all doped systems. 
Similar to the half-filling case~\cite{MJ2020}, the mechanism of AF enhancement can be traced to non-monotonic $S^f_{AF}(V)$ of homogeneous PAM shown in Fig.~\ref{Safden}(c). Obviously, $S^f_{AF}(V)$ is maximized at $V \sim 0.8-1.0t$ due to the competition between RKKY interaction mediated by $c-f$ hybridization $V$ and Kondo hybridization~\cite{AFnote}. Therefore, if $V_{1,2}$ approach to this critical $V$ range from two sides, the system can exhibit the AF enhancement in appropriate parameter regime. 

\begin{figure}[h!] 
\psfig{figure=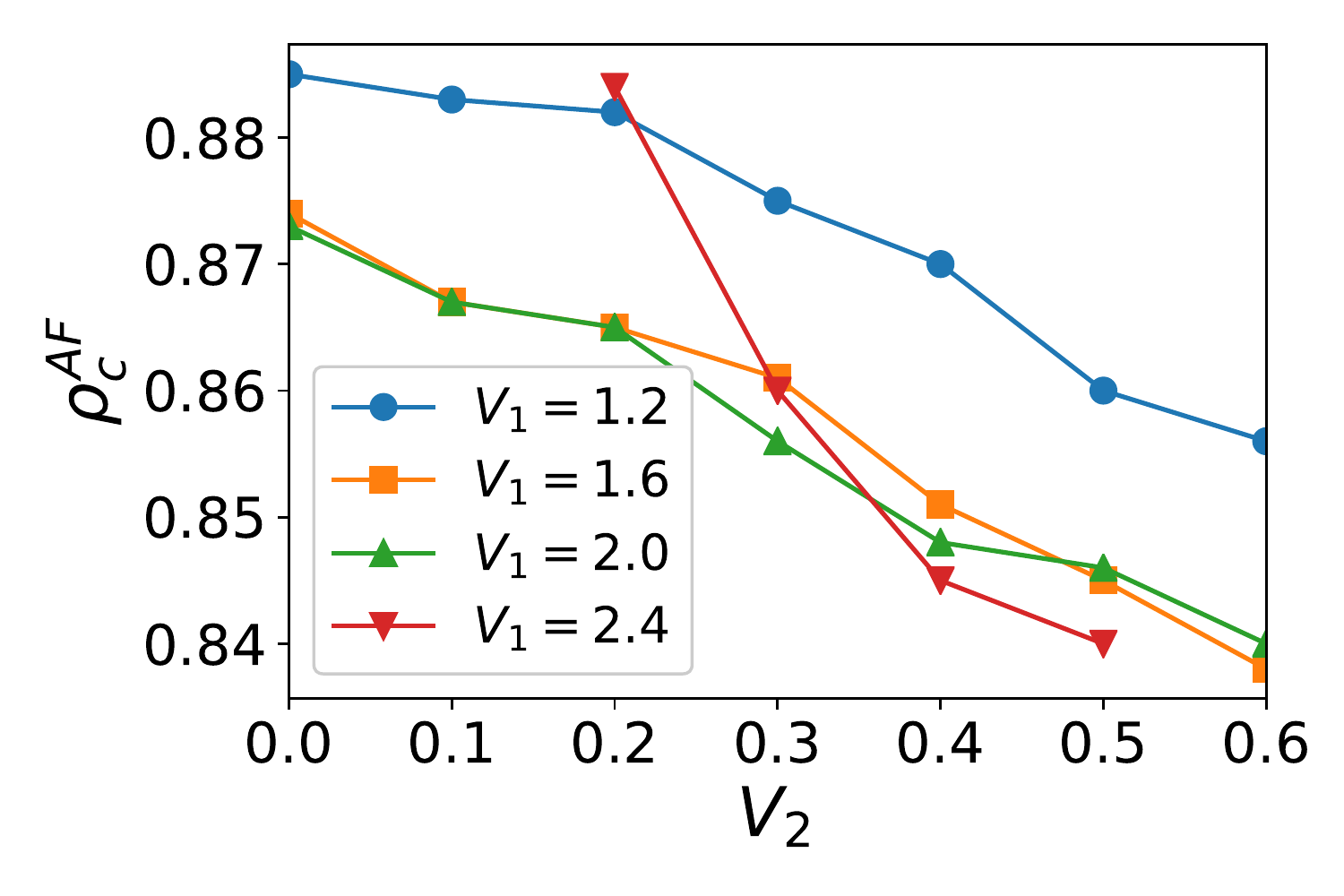, width=.5\textwidth,angle=0,clip}
\caption{(Color online) Critical density for the disappearance of AF enhancement as a function of $V_2$ for fixed $V_1$.}
\label{rhocAF}
\end{figure}  

Fig.~\ref{Safden}(a) has hinted the existence of a critical density $\rho^{AF}_c \sim 0.84$ below which the AF enhancement disappears. To further reveal this generic feature, Fig.~\ref{rhocAF} illustrates the evolution of the critical density versus $V_2$ for some fixed $V_1$. Apparently, $\rho^{AF}_c$ decreases with increasing $V_2$, which is consistent with the fact that the enhancement ratio at half-filling increases with $V_2$ so that the required doping level to destroy the enhancement is lowered accordingly.

\section{$d$-wave pairing enhancement}
\begin{figure}[h!] 
\psfig{figure=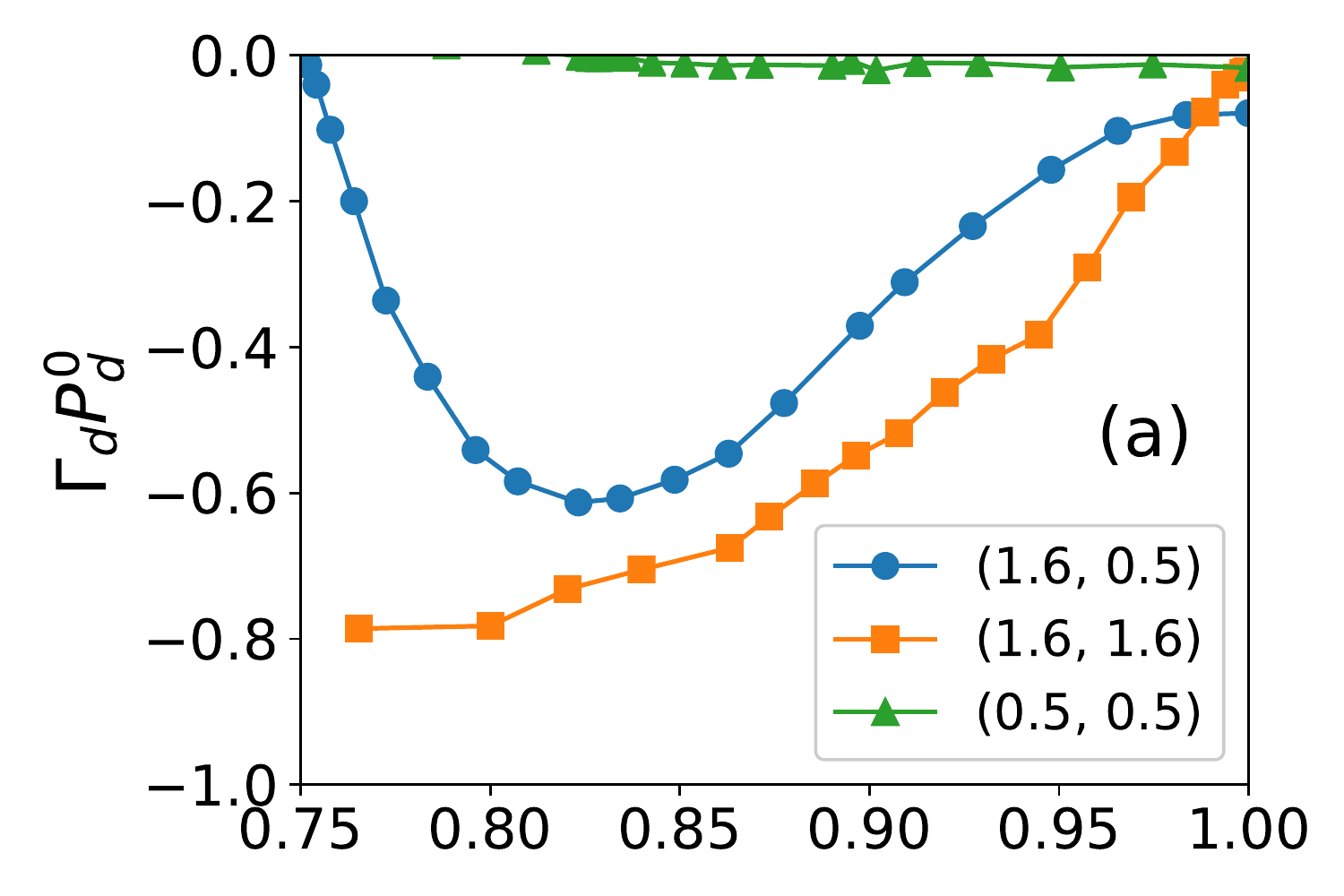,
width=.5\textwidth,angle=0,clip=true, trim = 0.0cm 0.5cm 0.0cm 0.2cm} \\
\psfig{figure=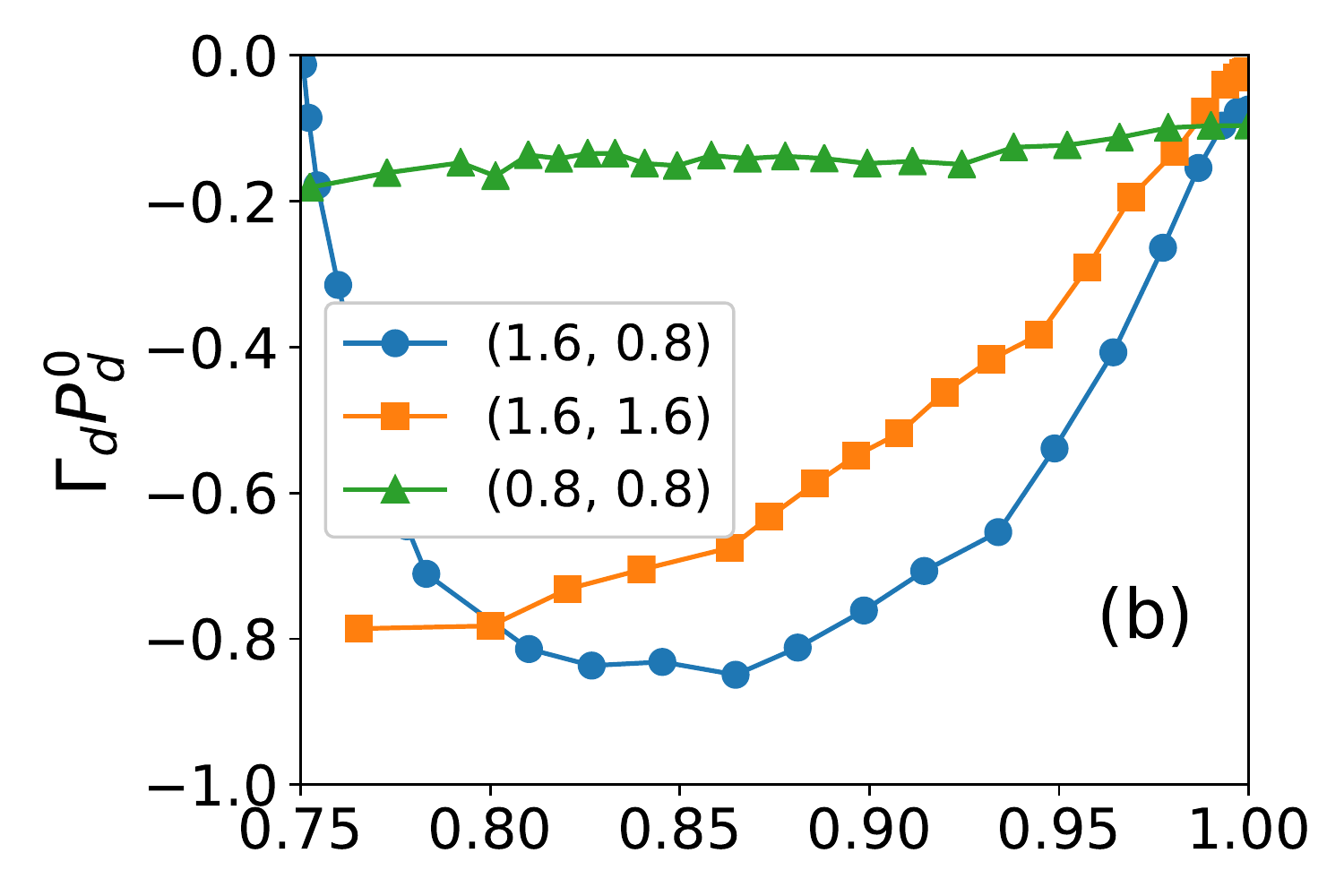,
width=.5\textwidth,angle=0,clip=true, trim = 0.0cm 0.5cm 0.0cm 0.2cm} \\
\psfig{figure=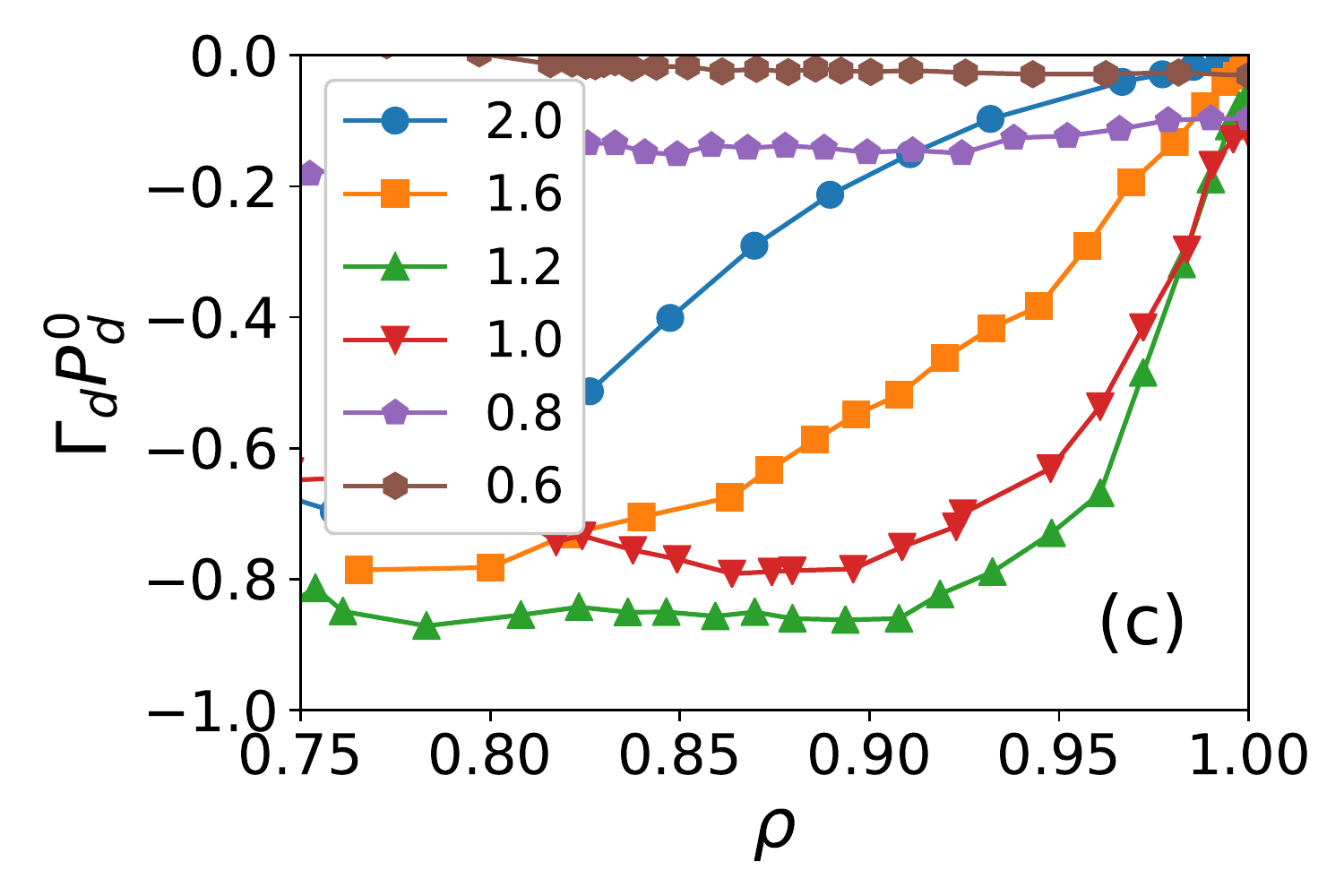,
width=.5\textwidth,angle=0,clip} 
\caption{(Color online) Doping evolution of $\Gamma_d P^0_d$ at $\beta t=15$ similar to Fig.~\ref{Safden}.}
\label{Pdden}
\end{figure}

As discussed before, one important feature of the doped conventional PAM is the competition and coexistence between the antiferromagnetism and $d$-wave superconductivity (dSC)~\cite{WeiWu}. Given that the striking AF enhancement can extend to a wide range of doping levels, whether the enhanced AF competes and/or coexists with dSC in the sPAM emerges as an intriguing question. 
To this aim, similar to Fig.~\ref{Safden}, Figure~\ref{Pdden}(a-b) illustrates the doping evolution of $\Gamma^{\phantom{0}}_d P^0_d$ of two characteristic systems with distinct behavior compared with their homogeneous counterparts. Fig.~\ref{Pdden}(c) provides the referenced evolution of $\Gamma^{\phantom{0}}_d P^0_d(V)$ in homogeneous PAM to facilitate the understanding the mechanism of dSC enhancement. 
There is no doubt that the half-filled system does not host dSC so that $\Gamma^{\phantom{0}}_d P^0_d$ is negligible. Doping the system gradually induces the dSC, whose diverging instability occurs at $\Gamma^{\phantom{0}}_d P^0_d=-1$. Note that in all systems with $V_1>V_2$, the dSC will disappear at $\rho=0.75$ because of the insulating nature of the ground state in the noninteracting limit, whose details can be found in Appendix.~\ref{U0}.

The most interesting feature is the dSC enhancement in a wide range of doping levels shown in Fig.~\ref{Pdden}(b), which again can be explained by the existence of the non-monotonic dependence of $\Gamma^{\phantom{0}}_d P^0_d$ upon varying $V$ in homogeneous PAM shown in Fig.~\ref{Pdden}(c), which demonstrates that the maximum dSC pairing tendency occurs at $V\sim1.2$ for all doping levels. Hence, if $V_{1,2}$ both exceed $\sim 1.2t$ or not, there is no dSC enhancement. Nonetheless, if $V_{1,2}$ approaches to $\sim 1.2t$ from two sides, a visible enhancement is possible in appropriate settings of $(V_1,V_2)$. We emphasize that if the AF enhancement is somewhat expected due to the enforced staggered pattern of $c-f$ hybridization, the enhancement of dSC is more fascinating because of the different form factor signs between a particular $f$ local moment and its four nearest neighbors for $d$-wave pairing.

\begin{figure}[h!] 
\psfig{figure=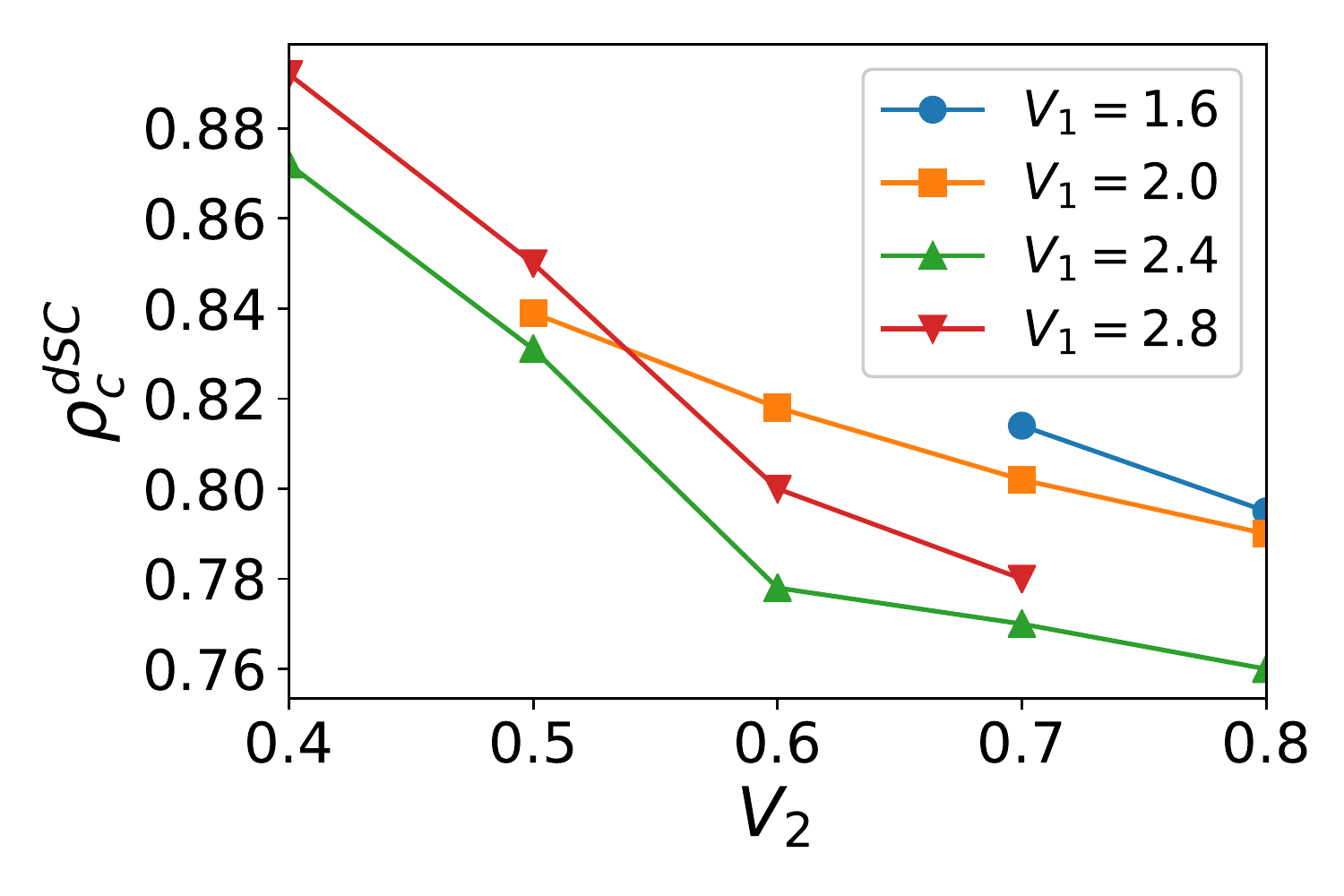, width=.5\textwidth,angle=0,clip}
\caption{(Color online) Critical density for disappearance of $d$-wave pairing enhancement as a function of $V_2$ for fixed $V_1$.}
\label{rhocPd}
\end{figure}

We have mentioned that the insulating ground state results in the vanishing of dSC at $\rho=0.75$, whose direct consequence is the disappearance of dSC enhancement at a critical density $\rho^{dSC}_c\sim 0.8$, for instance, shown in Fig.~\ref{Pdden}(b).  
More details of this generic feature is displayed in Fig.~\ref{rhocPd}, which illustrates the evolution of $\rho^{dSC}_c$ versus $V_2$ for some fixed $V_1$. Similar to $\rho^{AF}_c$ for the disappearance of AF enhancement, the critical density decreases with increasing $V_2$ because of the larger enhancement of $\Gamma^{\phantom{0}}_d P^0_d$ at smaller $V_2$.

\section{Phase diagram}
\begin{figure}[b] 
\psfig{figure=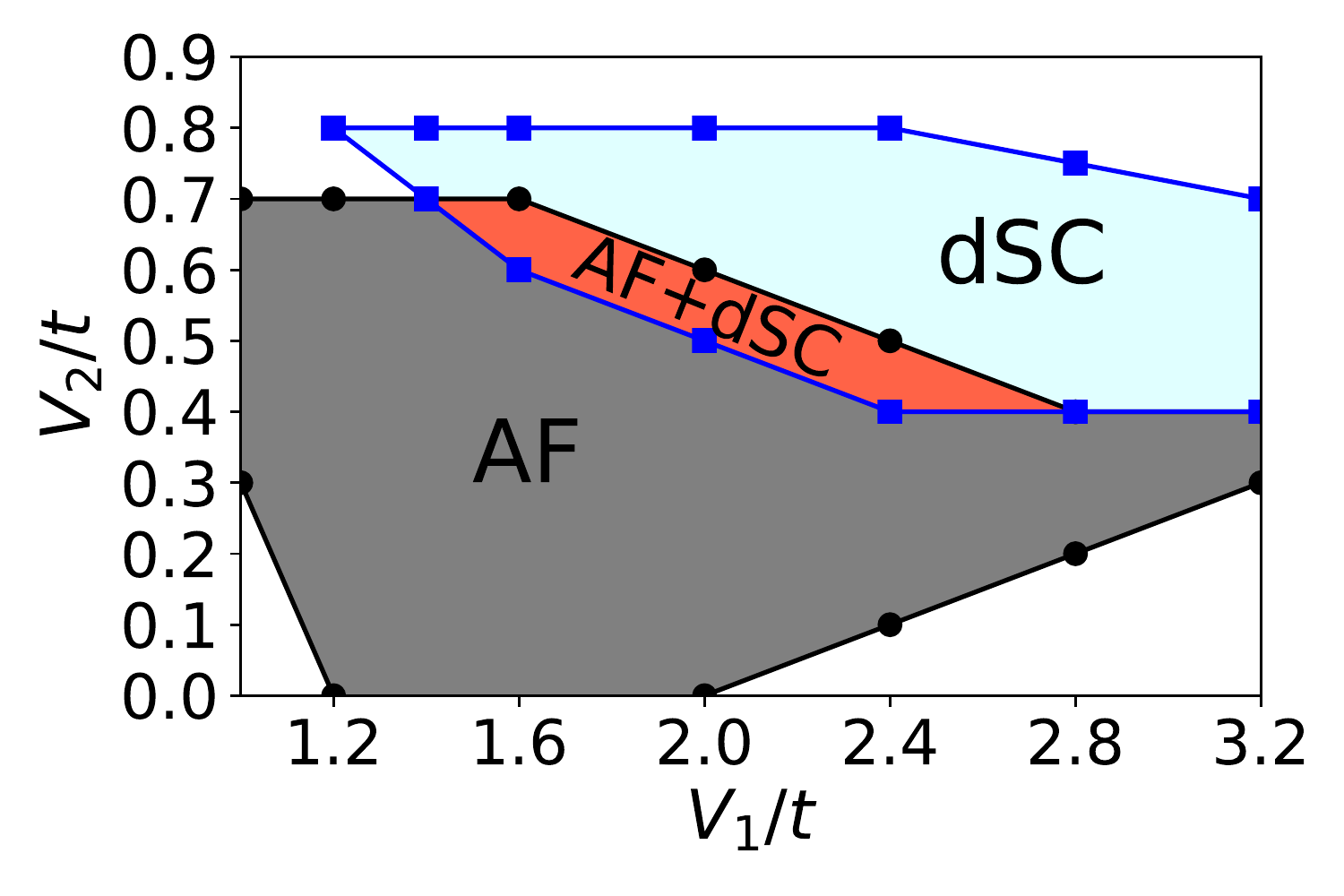}.pdf, height=6.2cm,width=.5\textwidth,angle=0,clip}
\caption{(Color online) Tentative phase diagram of staggered PAM for $\rho=0.95$ at lowest simulated temperature $T=0.05t$. The gray/cyan regime exhibits the enhancement of AF/dSC ordering tendency separately. The striking coexistence regime is highlighted by orange color. The white regime denotes the absence of the enhancements.}
\label{phase}
\end{figure}

Now that we have provided strong evidence of the enhanced ordering tendency of AF and dSC in sPAM individually, a significant question naturally arises regarding their competition and/or coexistence.
To this goal, we focus on a characteristic average density $\rho=0.95$~\cite{WeiWu}. Figure~\ref{phase} displays the tentative phase diagram at our lowest simulated temperature $T=0.05t$. Specifically, the gray/cyan regime exhibits the enhancement of AF/dSC separately, which vividly shows the competition between two enhancement. Note that the enhancement of dSC normally occurs at larger $V_2$ compared with that of AF, which can be understood as the consequence of the larger critical $V_2$ for the maximal dSC in homogeneous PAM (see Fig.~\ref{Safden}(c) and Fig.~\ref{Pdden}(c)).  
In addition, to our best knowledge, dSC enhancement only occurs when $V_1 \ge 1.2t$, namely located in the conventional heavy Fermi liquid regime of homogeneous PAM.

\begin{figure}[h!] 
\psfig{figure=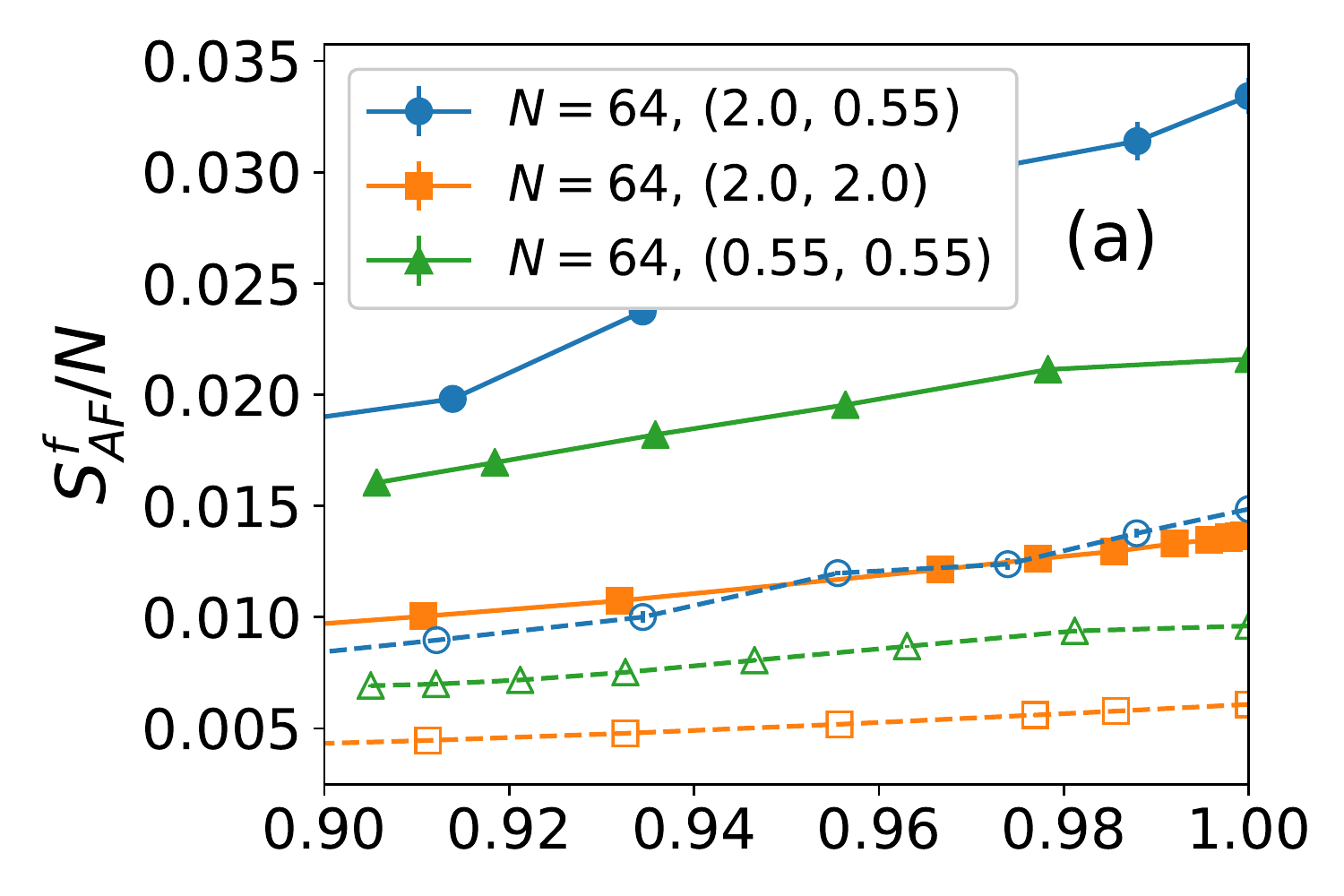,
width=.5\textwidth,angle=0,clip=true, trim = 0.0cm 0.5cm 0.0cm 0.2cm} 
\psfig{figure=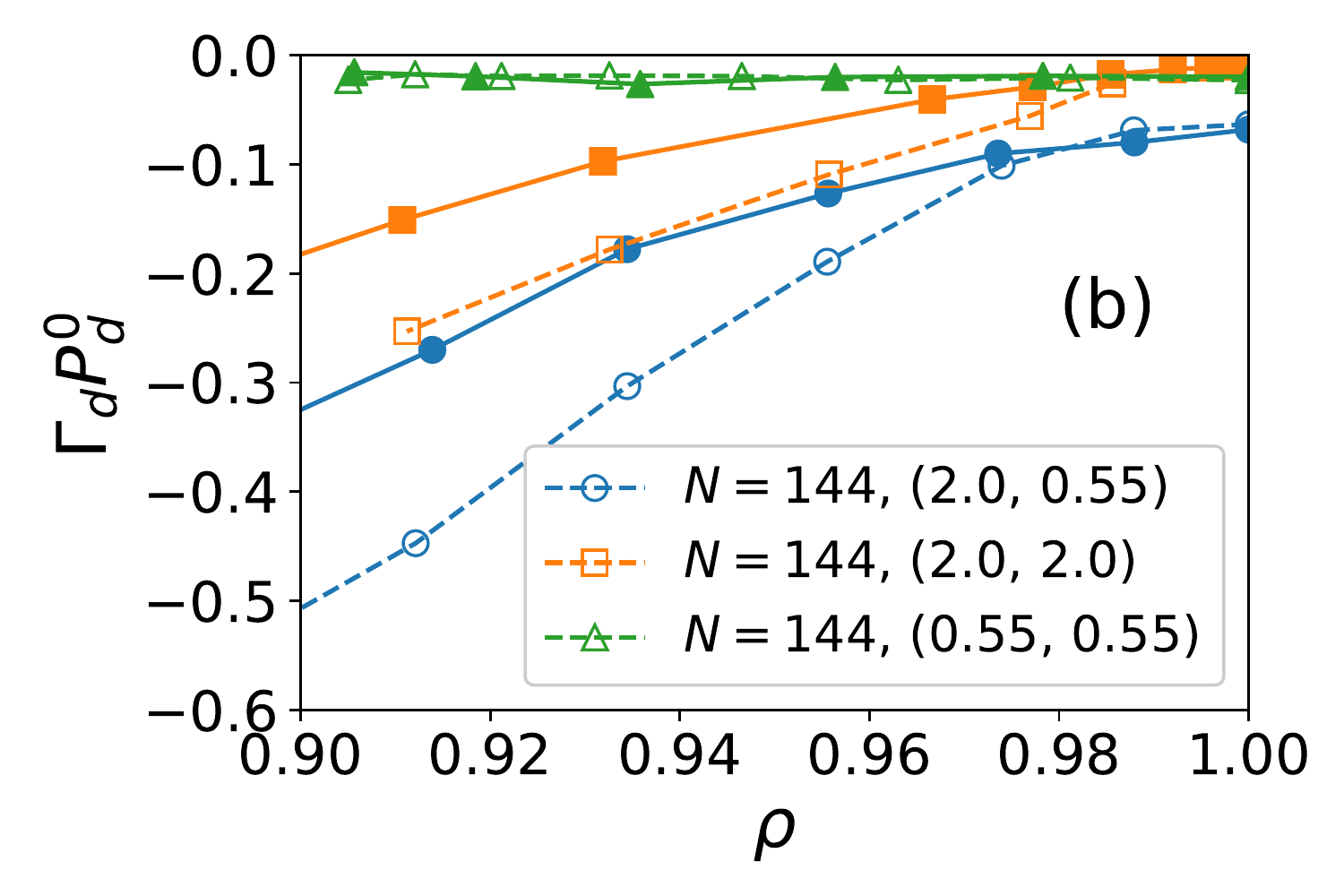,
width=.5\textwidth,angle=0,clip} 
\caption{(Color online) Doping evolution of (a) $S^f_{AF}/N$ and (b) $\Gamma_d P^0_d$ at characteristic $(V_1,V_2)=(2.0,0.55), \beta t=15$ of the coexistence regime in Fig.~\ref{phase}. The curves at a larger lattice $N=12\times12$ do not qualitatively modify the enhancement.}
\label{coexist}
\end{figure}

The most striking feature of the phase diagram is undoubtedly the coexistence regime highlighted by orange color, which can be treated as the crossover between the enhancement of two distinct physical quantities. Although the competition between AF and dSC is widely observed in quantum materials with strong electronic correlations, there is indeed strong theoretical~\cite{WeiWu} and experimental~\cite{ternary6} evidence to support their coexistence, especially in heavy fermion materials, due to unclear physical reasons until now. 
To further support our numerical findings of the narrow coexistence regime, Figure~\ref{coexist} demonstrates the doping evolution of (a) $S^f_{AF}/N$ and (b) $\Gamma^{\phantom{0}}_d P^0_d$ at characteristic $(V_1,V_2)=(2.0,0.55)$ of the coexistence regime in Fig.~\ref{phase}, which shows the ubiquity of the enhancement in a wide range of dopings. In addition, the robustness of their coexistence is further evidenced by the results at a larger lattice size $N=12\times12$, which do not qualitatively modify the general double enhancement. In all our simulations, in fact, the finite size effect is not significant for the occurrence of the AF and dSC enhancement. Although we are not able to provide a deterministic conclusion regarding the fate of the enhancement, especially the coexistence regime, at lower temperature scale, e.g. Neel temperature $T_N$ and $d$-wave superconducting transition $T_c$, our tentative phase diagram provides the guidance for future exploration of the enhancement of physical properties in sPAM and other generic systems. Last but not least, similar to the phase diagram shown in our previous work~\cite{MJ2020}, we emphasize that the enhancement regime of AF/dSC in Fig.~\ref{phase} precisely denote the enhanced ordering tendency but does not necessarily mean that the labeled (unlabeled white) regime can (cannot) definitely host the AF/dSC order, whose existence deserves more thorough investigation with sophisticated methods at lower temperature in the thermodynamic limit, which is out of the scope in this work.

\section{Conclusion}
In conclusion, we have provided strong numerical evidence of the remarkable enhancement of both the lattice antiferromagnetic structure factor and $d$-wave pairing in appropriate parameter regime via the determinant QMC simulations of the doped staggered PAM with two alternating inequivalent local moments. We demonstrate that the previously found enhanced AF ordering tendency at half-filling persists to a wide range of doping levels until a critical density below which the enhancement disappears. More interestingly, the $d$-wave pairing tendency can exhibit a similar enhancement in a range of doping levels due to the same physical mechanism.
Moreover, by providing a tentative phase diagram at a particular density, we demonstrated that the most striking feature lies in the coexistence of the enhanced tendency of both AF ordering and $d$-wave pairing, despite that its robustness against further lowering the temperature scale and relevance to realistic heavy fermion materials deserve further study.

To our best knowledge, these counterintuitive phenomena share the common thread of (a) the generic ``self-averaging'' effect between two distinct local moments and (b) the non-monotonic dependence of the corresponding physical quantity in homogeneous PAM. Therefore, this mechanism provides a new route of strengthening a particular ordering tendency in systems with inhomogeneity. If the above two essential ingredients turns out to be sufficient conditions, it is plausible to expect similar enhancement of physical properties in generic inhomogeneous systems, which deserves future investigation~\cite{note0}. In addition, it is requisite to explore the connection of our findings reported here to the realistic family of heavy fermion materials~\cite{ternary6,ternary7,Custers2020,ternary8} and their potential realization in ultracold atomic quantum simulator.

\begin{acknowledgments}
We acknowledge Mona Berciu and Richard Scalettar for useful discussion and the latter for providing the computing program of the non-interacting limit. This work was funded by the Stewart Blusson Quantum Matter Institute at University of British Columbia, and by the Natural Sciences and Engineering Research Council of Canada.
\end{acknowledgments}

\appendix
\section{Non-interacting band structure and density of states (DOS)}\label{U0}
In the absence of the inequivalence between two neighboring local moments, the conventional PAM in square lattice is recovered and the non-interacting limit at $U=0$ has two bands for each spin,
\begin{equation}
E_{1,2}(\mathbf{k})=\frac{1}{2} [ \epsilon_{\mathbf{k}} \pm \sqrt{\epsilon^2_{\mathbf{k}} + 4V^2} ] 
\label{nonintbands}
\end{equation}
where $\epsilon_{\mathbf{k} \sigma}=-2t(\cos k_{x}+\cos k_{y})$. 

The staggered PAM can be treated as a homogeneous PAM with average hybridization $\bar{V}_1 = \frac{1}{2}(V_1 + V_2)$ plus an additional staggered hybridization with amplitude $\bar{V}_2 = \frac{1}{2}(V_1 - V_2)$. In this picture, similar to the staggered potential in ionic Hubbard model~\cite{RTS2007,MJ2016}, incorporating a staggered hybridization on a bipartite lattice mixes the momentum states at ${\bf k}$ and ${\bf k + \pi}$ to form four energy bands and opens up additional spectral gaps at particular fillings. 

Therefore, we can rewrite the hamiltonian as
\begin{equation}
\begin{split}
{\cal H}_0 = 
\left(
\begin{array}{l}
  c^{\dagger}_{\mathbf{k}} \\
  c^{\dagger}_{\mathbf{k}+\pi} \\
  f^{\dagger}_{\mathbf{k}} \\
  f^{\dagger}_{\mathbf{k}+\pi} \\
\end{array}
\right)^{T}
\left(
\begin{array}{cccc}
  \epsilon_{\mathbf{k}}  & 0 & \bar{V}_1 & \bar{V}_2 \\
  0 & -\epsilon_{\mathbf{k}} & \bar{V}_2 & \bar{V}_1 \\
  \bar{V}_1 & \bar{V}_2 &  0  & 0 \\
  \bar{V}_2 & \bar{V}_1 &  0 & 0  \\
\end{array}
\right)
\left(
\begin{array}{l}
  c^{\phantom{\dagger}}_{\mathbf{k}} \\
  c^{\phantom{\dagger}}_{\mathbf{k}+\pi} \\
  f^{\phantom{\dagger}}_{\mathbf{k}} \\
  f^{\phantom{\dagger}}_{\mathbf{k}+\pi} \\
\end{array}
\right)
\end{split}
\label{Hk}
\end{equation}
where $\bar{V}_{1,2} = \frac{1}{2}(V_1 \pm V_2)$ so that the four-band structure is given by $E^2(\mathbf{k})=\frac{1}{2} (A \pm \sqrt{A^2-4V^2_1 V^2_2})$ with $A \equiv \epsilon^2_{\mathbf{k}}+V^2_1+V^2_2$ as the generalization of homogeneous PAM.

\begin{figure}[h!] 
\psfig{figure=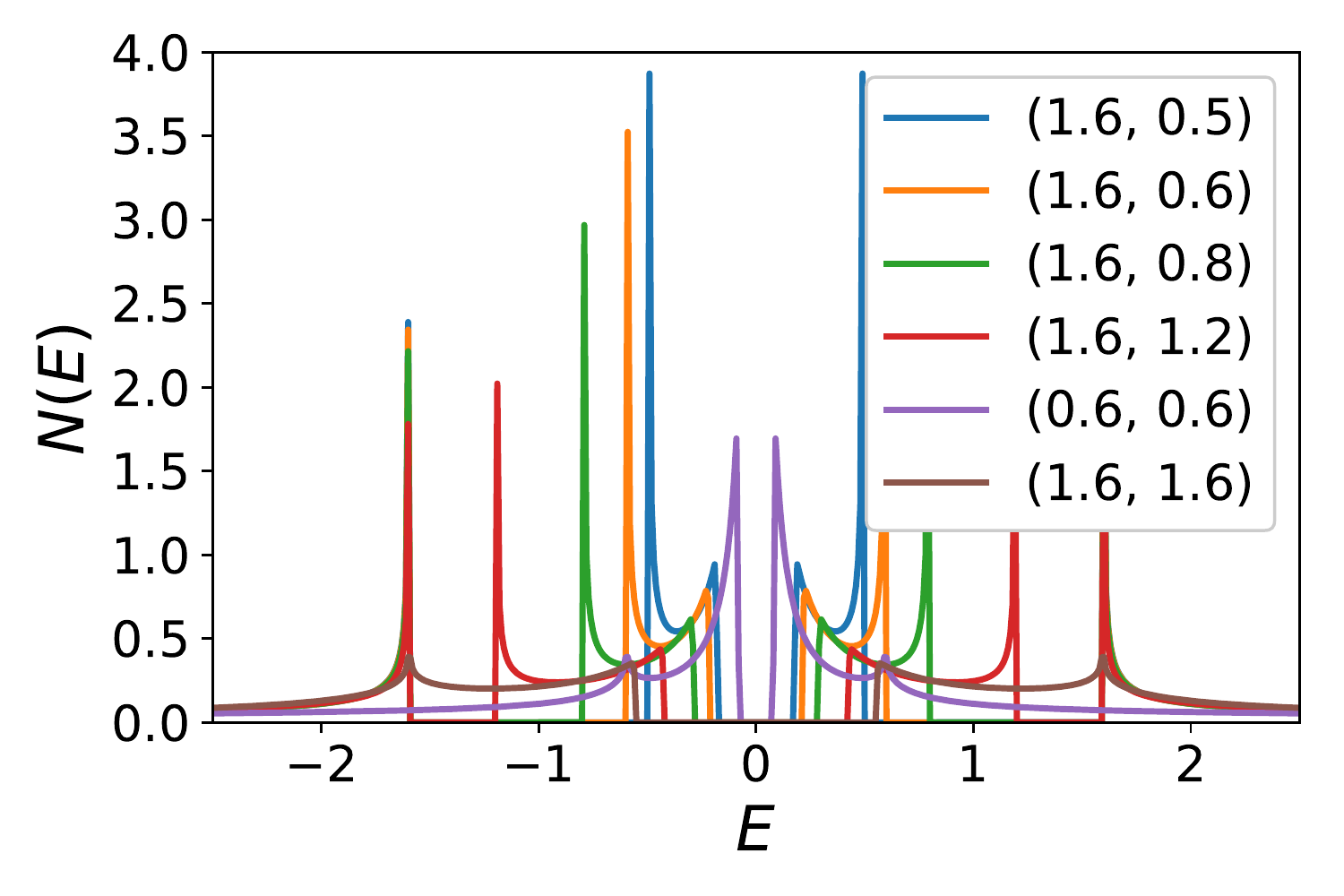, width=.5\textwidth,angle=0,clip}
\caption{(Color online) Density of state at $U=0$ of some characteristic cases of staggered and homogeneous PAM.}
\label{dosU0}
\end{figure}

More information can be gathered by the density of states shown in Figure~\ref{dosU0}. It is tempting to explore whether the enhancement of AF and dSC reported in the main text can be induced by some obvious effects of the DOS at $U=0$, which might
be enhanced by the inhomogeneity of hybridization in sPAM that in turn leads to a stronger RKKY interaction or so. Nonetheless, Fig.~\ref{dosU0} does not support this physical picture. For instance, although the $(V_1,V_2)=(1.6,0.6)$ system exhibits the AF enhancement while the $(V_1,V_2)=(1.6,1.2)$ counterpart does not, as shown in Fig.~\ref{Safden}, their corresponding DOS at $U=0$ do not show significant difference except the expected shift of spectral weights.
Furthermore, neither the hybridization gap around $E=0$ nor the total bandwidth has strong implication on whether the AF and/or dSC enhancement can appear.
One important feature is new hybridization gaps opening at the 1/4 and 3/4 fillings, which explains the vanishing of $\Gamma^{\phantom{0}}_d P^0_d$ in Fig.~\ref{Pdden}.

\section{DQMC and fermionic sign problem}\label{dqmc}
To treat with the Hubbard interaction together with diverse energy scales associated with two inequivalent local moments on the equal footing, we solve Eq.~\ref{inPAM} by means of the finite temperature determinant Quantum Monte Carlo (DQMC)~\cite{blankenbecler81}, where a path integral expression is written for the quantum partition function ${\cal Z}={\rm Tr\,exp}\,(\,-\beta {\cal H} \,)$. The Hubbard interaction term is rewritten as a coupling of the electron spin with a space and imaginary-time dependent auxiliary field so that the fermionic degrees of freedom can be integrated out analytically with the payment of the additional auxiliary field. The integrals over the field configurations are sampled in a Monte Carlo algorithm. Despite that the half-filled case~\cite{MJ2020} of Eq.~\ref{inPAM} ensures the absence of infamous Fermionic sign problem so that the physical quantities can be evaluated at low enough temperatures~\cite{loh90}, the doped system possesses more difficulty due to the sign problem, which strongly depends on the doping level, lattice size, and temperature scale etc. 

\begin{figure}[h!] 
\psfig{figure=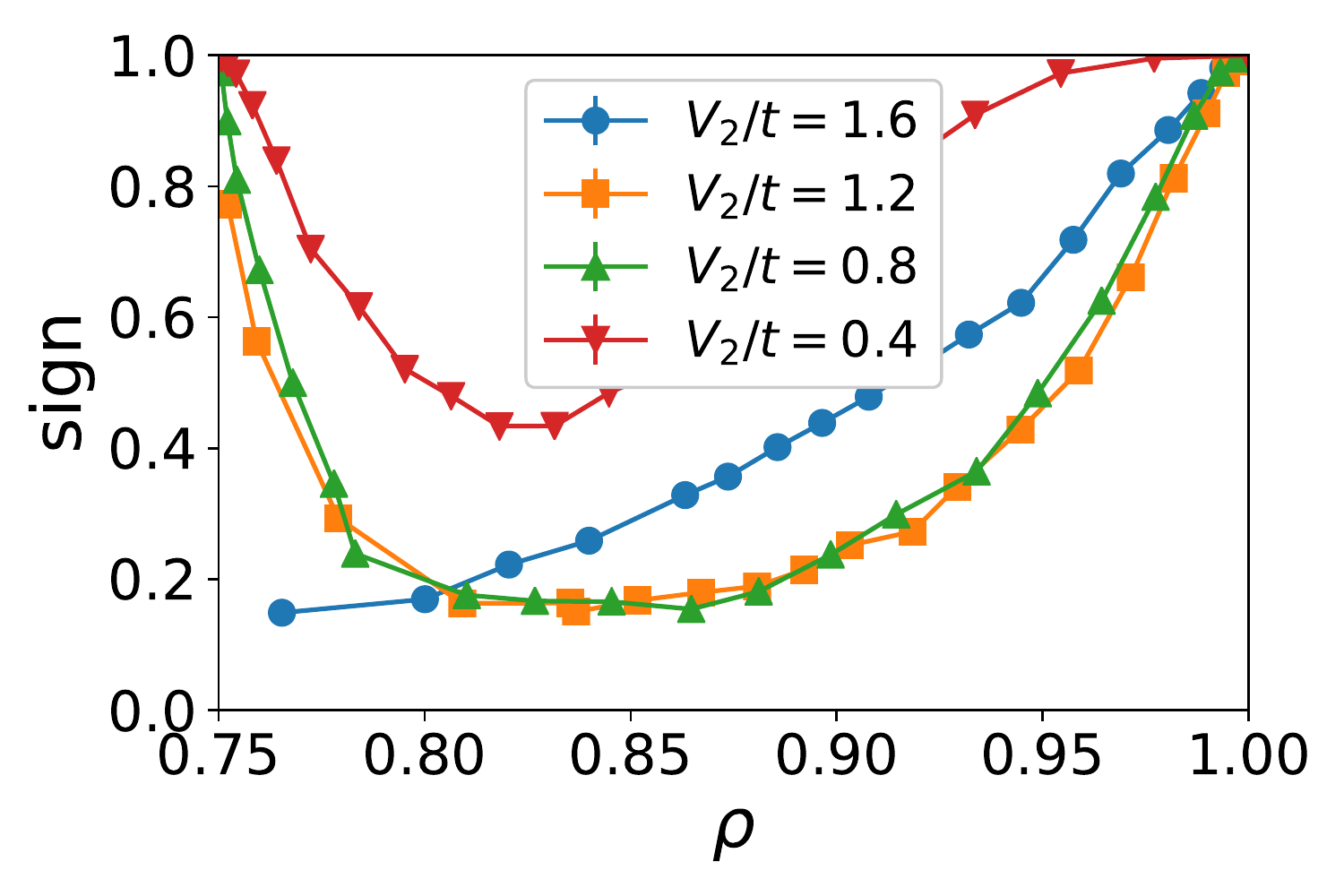, width=.5\textwidth,angle=0,clip}
\caption{(Color online) Representative evolution of QMC sign problem versus the average density at $V_1/t=1.6, \beta t=15, N=8 \times 8$. The larger lattice sizes, e.g. $N=12 \times 12$ have more severe sign problems in general.}
\label{sign}
\end{figure}

Figure.~\ref{sign} provides a representative example of the behavior of the DQMC sign problem versus doping. Although there is no general theories to predict, interpret, and tackle with the sign problem, some specific features can be seen in the case of sPAM. Interestingly, the sign problem shows a non-monotonic dependence of $V_2$ at fixed $V_1$. In addition, the dominant difference between the homogeneous and sPAM lies in the vanish of the sign problem of the latter at $\rho=0.75$. This is consistent with the general expectation that the insulating systems normally have much better performance of the fermionic sign problem.

\section{Charge structure factor}\label{cdw}
In the main text, we have focussed on the magnetic and superconducting pairing properties. Although the charge degree of freedom is frozen in the half-filled KLM/PAM, it can play an important role in the doped systems. In fact, the charge order in the two-dimensional KLM has been uncovered even in absence of the bare repulsive interactions~\cite{cdw}. In our sPAM, the $c$-$f$ hybridization has intrinsic checkerboard pattern, which naturally results in the corresponding charge density modulation.  

To explore the charge order and specifically whether it has similar enhancement in sPAM, we calculate the charge structure factor of both $c$ and $f$ orbitals 
\begin{equation}
S^{\alpha}_{CDW}(V_1,V_2)=\frac{1}{N} \sum_{ij} e^{-i \mathbf{q} \cdot (\mathbf{R_i}-\mathbf{R_j})} \langle (n^{\alpha}_{i\uparrow}+n^{\alpha}_{i\downarrow}) (n^{\alpha}_{j\uparrow}+n^{\alpha}_{j\downarrow}) \rangle
\end{equation}
at $\mathbf{q}=(\pi,\pi)$ with the orbital index $\alpha=c,f$, where $\mathbf{R_i}$ denotes the coordinates of site $i$ and $N$ is the lattice size. 

\begin{figure}[h!] 
\psfig{figure=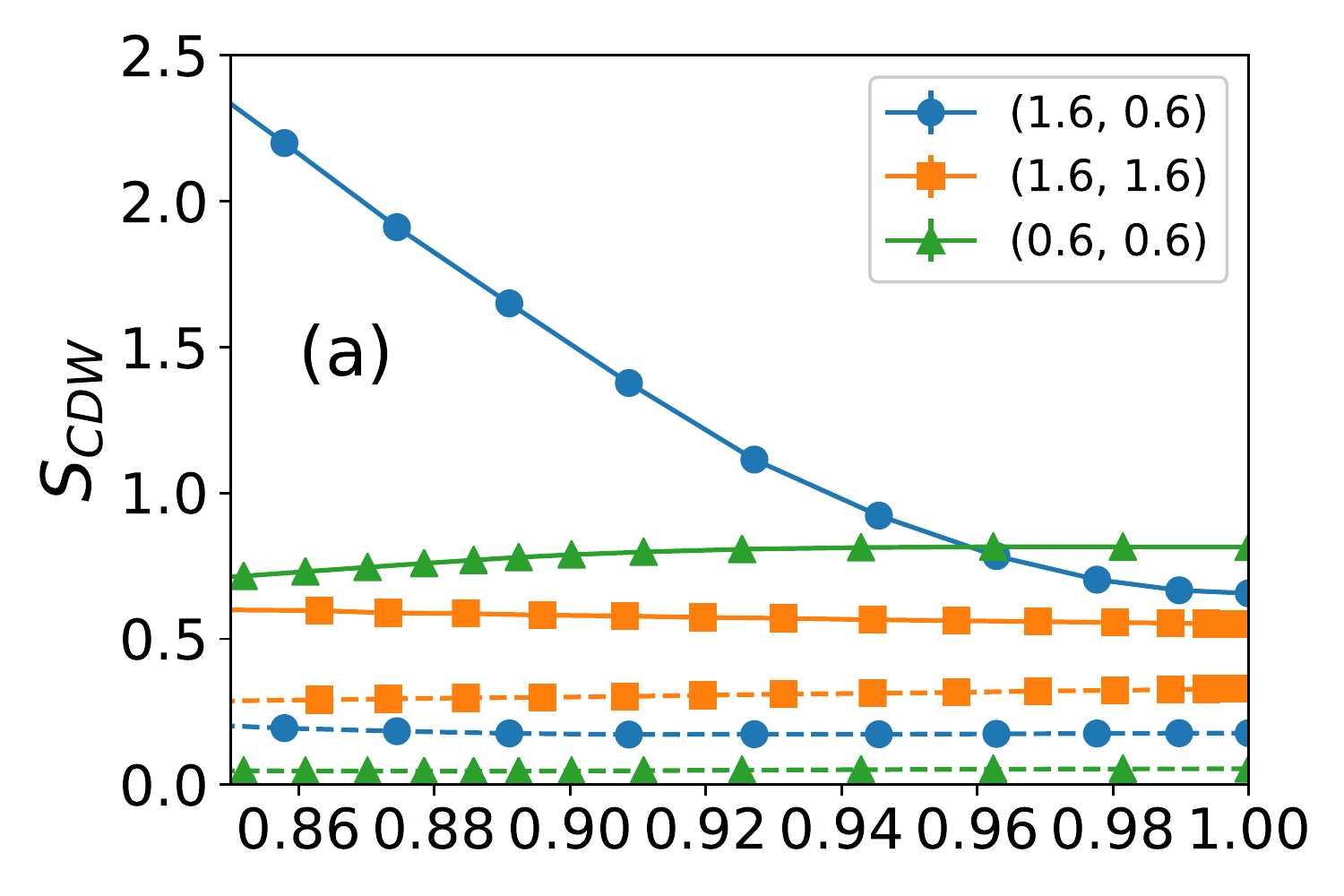,
width=.5\textwidth,angle=0,clip=true, trim = 0.0cm 0.5cm 0.0cm 0.2cm} \\
\psfig{figure=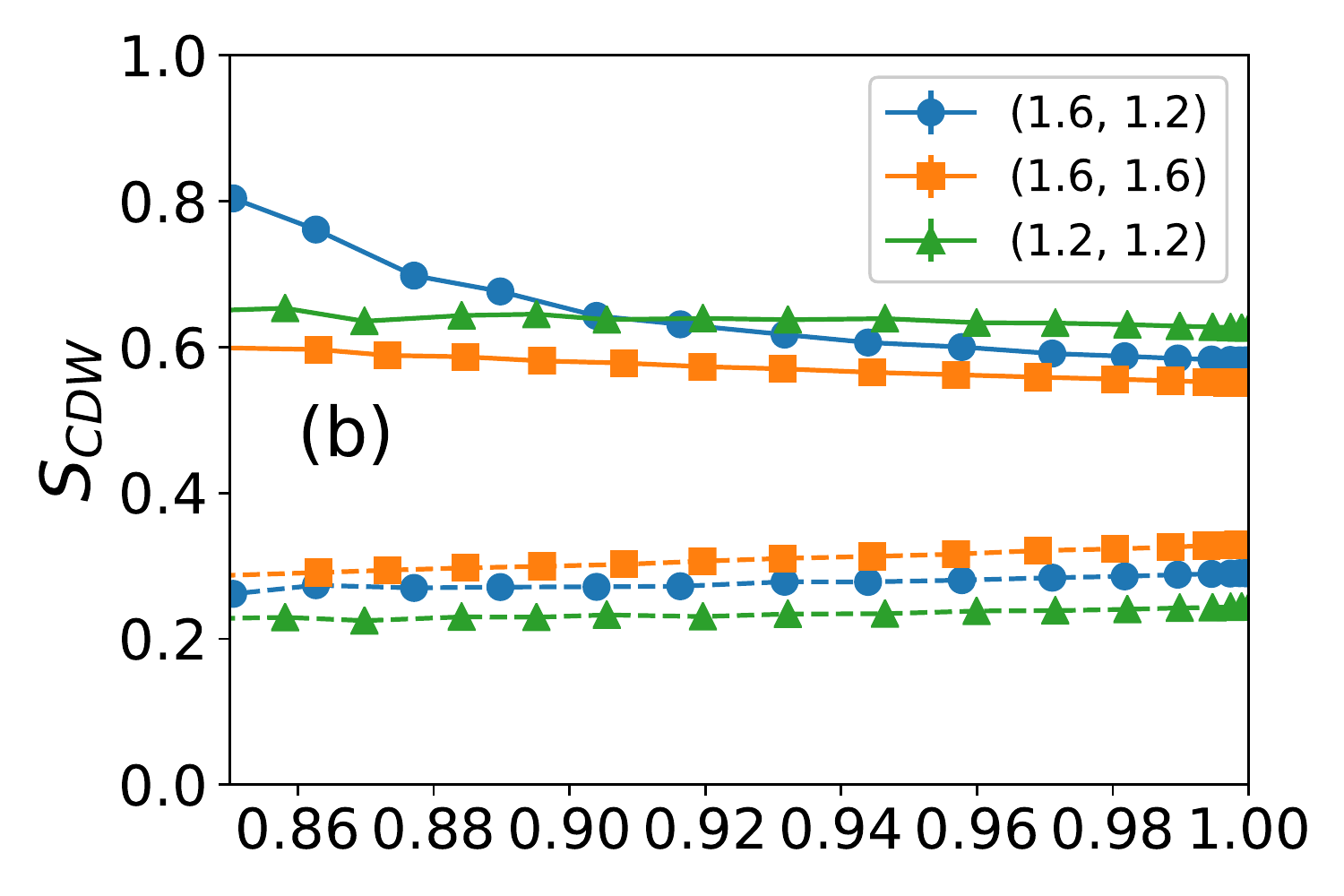,
width=.5\textwidth,angle=0,clip=true, trim = 0.0cm 0.5cm 0.0cm 0.2cm} \\
\psfig{figure=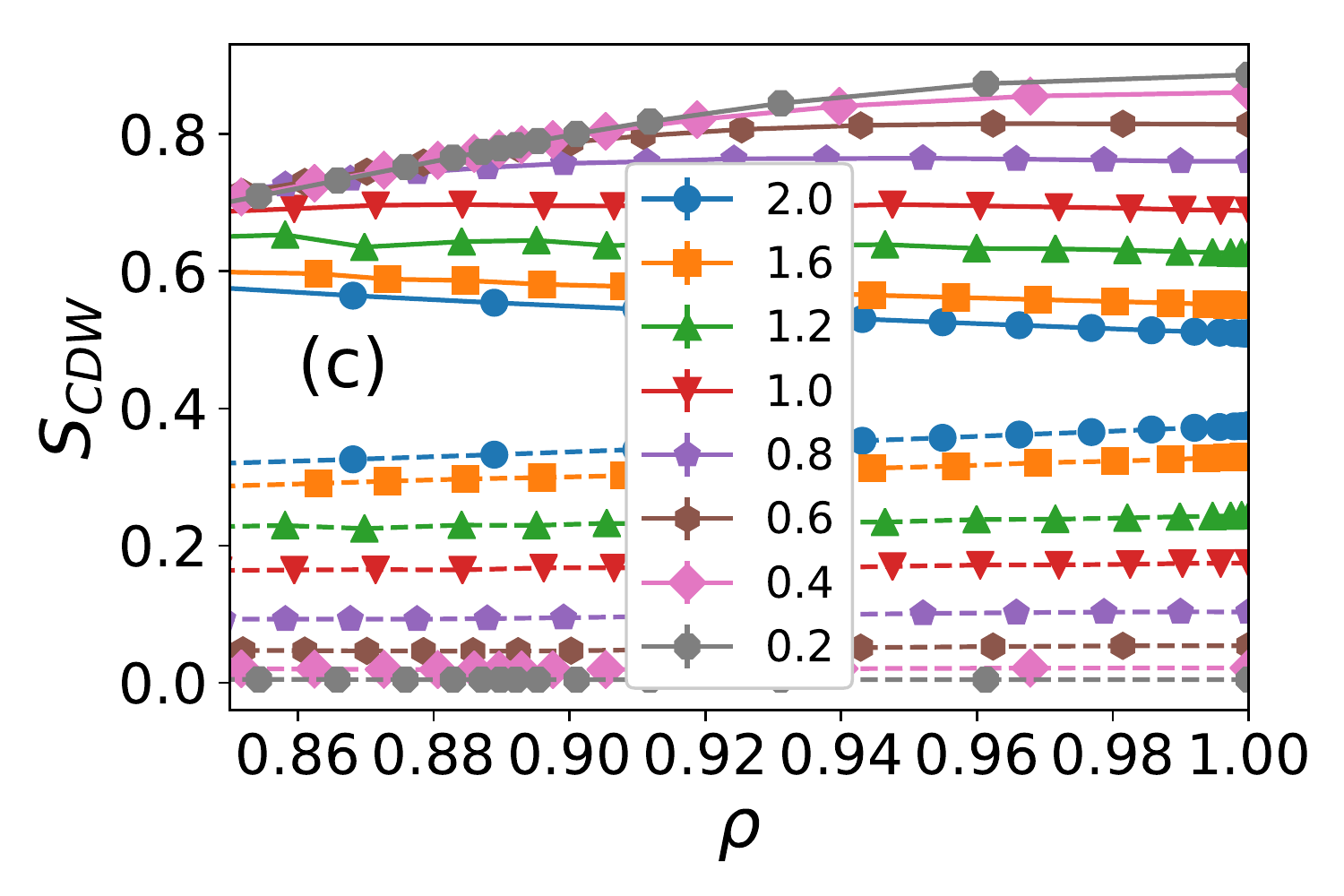,
width=.5\textwidth,angle=0,clip} 
\caption{Doping evolution of $S_{CDW}(V_1,V_2)$ at $N=8\times 8, \beta t=15$ in (a-b) two characteristic systems compared with their counterparts in homogeneous PAM and (c) homogeneous PAM. The legend is shown as $(V_1,V_2)$ of sPAM in (a-b) and $V$ of PAM in (c). The solid/dashed lines are for $\alpha=c, f$.}
\label{Scdw}
\end{figure}

To compare with the magnetic properties shown in Fig.~\ref{Safden}, Figure~\ref{Scdw}(a-b) show the doping evolution of $S^{\alpha}_{CDW}(V_1,V_2)$ of two characteristic systems compared with their homogeneous counterparts; Fig.~\ref{Scdw}(c) provides the referenced evolution of $S^{\alpha}_{CDW}(V)$ in homogeneous PAM. The solid/dashed lines are for $\alpha=c, f$ respectively. 

As expected, $c$ orbital has larger charge modulation due to the strong correlation of $f$-orbital. 
At half-filling $\rho=1$, in stark contrast to the AF enhancement in Fig.~\ref{Safden}(a), $S^{\alpha}_{CDW}(1.6,0.6)$ does not possess any enhancement. This can be understood by the behavior of homogeneous PAM in Fig.~\ref{Scdw}(c), where the absence of the non-monotonic dependence of $S^{\alpha}_{CDW}(V)$ ensures that the generic ``self-averaging'' effect results in $S^{\alpha}_{CDW}(V_1)<S^{\alpha}_{CDW}(V_1,V_2)<S^{\alpha}_{CDW}(V_2)$ in all cases.

Naturally, the doping introduces additional charges on $c$ orbital so that $S^{c}_{CDW}(V_1,V_2)$ grows quickly owing to the intrinsic checkerboard pattern of $c-f$ hybridization and finally exceeds its homogeneous counterparts. Distinct from $S^{f}_{AF}$, however, this is a generic feature for all combinations of $(V_1,V_2)$. Besides, there is no doubt that $S^{c}_{CDW}(V_1,V_2)$ increases if $V_1$ departs away from $V_2$ by comparing (a) and (b) panels.

Therefore, there is no enhancement of charge order due to the nontrivial non-monotonic dependence of $S^{\alpha}_{CDW}(V)$ in homogeneous PAM. Instead, the trivial enhancement of $S^{c}_{CDW}(V_1,V_2)$ in heavily doped sPAM is simply induced by the intrinsic checkerboard doping pattern on conduction bands. This further supports our major conclusion that the non-monotonic dependence is crucial to realize the nontrivial enhanced tendency of physical quantities like AF order and $d$-wave pairing.



\begin{thebibliography}{27}
\bibitem{note0}
Our preliminary results showed that the Hubbard model with staggered interaction strengths can host the enhanced AF structure factor as well, whose further confirmation is desired.
\bibitem{AFnote}
Note that the exact critical hybridization is still under debate. Ref.~\cite{pam5} recently reported that the quantum critical hybridization separating zero temperature antiferromagnetic insulator and a Kondo insulator of the two-dimensional PAM lies at $V_c/t=0.91$; while Ref.~\cite{wenjian} concludes that $V_c/t=1.0$.
\end{thebibliography}
\end{document}